\documentstyle[twocolumn,pra,aps]{revtex}
\draft
\newcommand{\be}{\begin{eqnarray*}}
\newcommand{\ee}{\end{eqnarray*}}
\newcommand{\sss}{\scriptscriptstyle}
\begin{document}
\twocolumn[\hsize\textwidth\columnwidth\hsize\csname 
@twocolumnfalse\endcsname
\title{Reduced phase space quantization}

\author{Pravabati Chingangbam\thanks{e-mail:prava@arbornet.org} 
and 
	Pankaj Sharan\thanks{e-mail:pankj.ph@jmi.ernet.in}}
\address{Department of Physics, Jamia Millia Islamia  \\
	New Delhi-110 025, India    }

\maketitle

\begin{abstract}
We examine two singular Lagrangian systems with constraints which 
apparently reduce the phase space to a 2-dimensional sphere and
a 2-dimensional hyperboloid. Rigorous 
constraint analysis by Dirac's method, however, gives 
2-dimensional open disc and an infinite plane with a hole in the 
centre 
respectively as the reduced phase spaces. Upon canonical 
quantisation 
the classical constraints show up as restrictions on the Hilbert space.
\end{abstract}

]

\section{Introduction}
In the study of quantization of classical systems one must start with
two essential things, namely, a phase space for the system and 
a dynamical principle. This principle may be a classical Hamiltonian 
derived from a Lagrangian or from a set of hyperbolic field 
equations. 
It may also be some quantum requirement such as annihilation of 
unphysical states by constraint operators.
In an interesting paper\cite{rv} Radhika Vathsan considered 
quantization of 
a\linebreak 4-dimensional phase space with canonical coordinates 
$\,(q^{\sss 1}, q^{\sss 2}, p_{\sss 1}, p_{\sss 2})\,$ with {\it a priori} 
chosen constraints
\be
\phi &\equiv& q^{{\sss 1}2} + q^{{\sss 2}2} + p_{\sss 1}^2 + p_{\sss 
2}^2 
                             - R^2  =0  \\
\chi &\equiv& p_{\sss 2} =0
\ee
using geometric and Dirac method of quantization. In her analysis
these constraints do not
follow from any Lagrangian. $\phi$ is arbitrarily assumed to be 
a first class constraint and $\chi$ is chosen as a gauge fixing 
condition. 
The reduced phase space turns out to be 2-dimensional sphere 
$S^2$ of radius $R/2$.

\noindent In the present paper we reanalyse quantization of the 
above system
using Dirac's method\cite{dirac,8,9,10}. 
Dirac's method starts with a singular Lagrangian which inherently 
contains the constraints. Whether the constraints are first or second 
class
follows in a straightforward manner from the analysis without any 
arbitrariness.
We choose here two Lagrangians, the first of which reproduces the 
same set of
constraints as\cite{rv} as a pair of second class constraints. 
The second example gives similar looking constraints but with a 
minus sign for the $(q^{\sss 2})^2$ term.
We do rigorous constraint analysis and then quantize canonically.

\section{Constraint analysis and quantization}

\noindent Consider the Lagrangian
$$ L= {{{\dot q}^{{\sss 1}2}}\over {4q^{\sss 2}}} - q^{\sss 2}
                            \Big( q^{{\sss 1}2} + {{q^{{\sss 2}2}}\over{3}} 
             - R^2 \Big)     \eqno(1)         $$
excluding the line $q^{\sss 2} = 0$ on the configuration space.	     
We solve for canonically conjugate momenta to get
\be
p_{\sss 1} &=& {{{\dot q}^{\sss 1}}\over{2q^{\sss 2}} }   \\
p_2 &=& 0
\ee
The second equation is a primary constraint
$$ \phi_1\equiv p_{\sss 2}=0                             \eqno(2) $$
The Hamiltonian is given by
$$ H = q^{\sss 2} p_{\sss 1}^2 + p_{\sss 2}v_{\sss 2} 
            + q^{\sss 2}\Big( q^{{\sss 1}2} + {{q^{{\sss 2}2}}\over{3}} 
             - R^2 \Big)                      \eqno(3) $$
where $v_{\sss 2}$ is unknown Lagrange multiplier.
By evolving $\phi_1$ and setting it to zero 
$$ \big\{p_{\sss 2},H\big\}=0  $$
we get a secondary constraint
$$ \phi_2\equiv q^{{\sss 1}2} + q^{{\sss 2}2} + p_{\sss 1}^2 + 
p_{\sss 2}^2 - R^2  =0   \eqno(4)   $$
Further evolution of $\phi_2$ determines $v_{\sss 2}$
$$v_{\sss 2}=0                                  \eqno(5)              $$

There are no further constraints. We,therefore, obtain two 
constraints 
$\phi_1,\,\phi_2$ with non-zero Poisson bracket between them and 
so are  
second class constraints. To get the reduced phase space the extra
degrees of freedom corresponding to 
these constraints must be completely removed. The 
Dirac bracket is defined by
$$ \big\{ f,g \big\}_D = \big\{ f,g \big\} - \big\{ f,\phi_i \big\}
                      \big( C^{-1} \big)_{ij} \big\{\phi_j,g \big\}  \eqno(6)$$
for any two classical observables $ f(q,p) ,\ g(q,p)$. The matrix $C$ 
is 
$$ C =
\left(\begin{array}{cc} 
          0     &  -2q^{\sss 2}   \\
          2q^{\sss 2}   &    0  
\end{array}
\right)	 
$$
where 
$$ C_{ij}= \big\{ \phi_i,\phi_j \big\} $$
The basic Dirac brackets are
\be
\big\{ q^{\sss 1},p_{\sss 1} \big\}_D  &=& 1  \\
\big\{ q^{\sss 2},p_{\sss 1} \big\}_D  &=& -{{q^{\sss 1}}\over{q^{\sss 
2}}}  \\
\big\{ q^{\sss 1},q^{\sss 2} \big\}_D  &=& -{{p_{\sss 1}}\over{q^{\sss 
2}}}  
\ee
$$\qquad\qquad\qquad\qquad\qquad\qquad                        
\eqno(7) $$
The rest are zero.

\noindent Next we put both constraints equal to zero. $\phi_1=0$ 
eliminates 
$p_{\sss 2}$ and from $\phi_2=0$ we eliminate $q^{\sss 2}$.
$$ q^{\sss 2} = \pm{\sqrt{R^2-p_{\sss 1}^2-q^{{\sss 1}2}}}      
\eqno(8) $$
There is $\pm $ sign ambiguity in $q^{\sss 2}$ which corresponds to 
the fact that the configuration space we started with consisted of 
two disconnected parts given by $ q^{\sss 2} > 0 $ and $ q^{\sss 
2}<0 $. 
This amounts to residual freedom in  obtaining the phase space 
even after
imposing all constraint conditions. For each choice of sign for $ 
q^{\sss 2}  $
eqn(8) gives
$$ p_{\sss 1}^2 + q^{{\sss 1}2} =  R^2 - q^{{\sss 2}2}     \eqno(9)   
$$
or,
$$ p_{\sss 1}^2 + q^{{\sss 1}2} <  R^2                     \eqno(10)   $$
The completely reduced phase space is, therefore, a disc of radius 
$R$   
without the boundary. 
Choosing $q^{\sss 2}$ to be positive the reduced Hamiltonian is
$$ H= {{2}\over{3}} \Big( R^2-p_{\sss 1}^2-q^{{\sss 1}2} \Big)^{3/2}  
\eqno(11)	$$
The equations of motion are 
\be
{\dot q}^{\sss 1} = \big\{ q^{\sss 1},H \big\}  &=& {{\partial H}\over 
{\partial p_{\sss 1}}} \\	      
                              {}  &=& 2q^{\sss 2}p_{\sss 1}      \\            
{\dot p}_{\sss 1} = \big\{ p_{\sss 1},H \big\}  &=& -{{\partial H}\over 
{\partial q^{\sss 1}}} \\
                              {}  &=& -2q^{\sss 2} q^{\sss 1}        
\ee			       
$$ \qquad\qquad\qquad\qquad\qquad\qquad                       
\eqno(12)   $$

\noindent We proceed to canonically quantise the system. Hilbert 
space 
${\mathcal H}$ consists of configuration space wave-functions 
$\,\psi(q)$ 
(index 1 has been dropped from $q^{\sss 1}$) which are square 
integrable
in the interval $-R<q<R$.
Observables are self-adjoint operators on ${\mathcal H}$. 
$\,q$ and $p$ go over to position and momentum operators 
$$ q^{\sss 1} \rightarrow {\hat q}  $$
$$ {\hat q}\,\psi(q)=q\,\psi(q)                            \eqno(13)    $$
$$ p_1\rightarrow {\hat p} \equiv -i\hbar{{\partial}\over{\partial q}}
                                                            \eqno(14)   $$
and ${\hat q},{\hat p}$ satisfy the commutation relation
$$ \big[ {\hat q}, {\hat p}\big]=i\hbar                    \eqno(15)    $$
The evolution of the system is generated by the Hamiltonian operator
$$ {\hat H}={{2}\over{3}} \Big( R^2 - {\hat p}^2 -{\hat q}^2
                                \Big)^{3/2}         \eqno(16)    $$
which must be self-adjoint for the evolution to be unitary.

Further, the operator 			     
$ \,\big( R^2 - {\hat p}^2 - {\hat q}^2 \big)\,$ must have
positive eigenvalues for ${\hat H}$ to be positive-definite. 
This requirement is due to the classical constraints showing up
at the quantum level.

Consider next the Lagrangian 
$$ L= {{{\dot q}^{{\sss 1}2}}\over {4q^{\sss 2}}} - q^{\sss 2}
                            \Big( q^{{\sss 1}2} - {{q^{{\sss 2}2}}\over{3}} 
             - R^2 \Big)                             \eqno(17)    $$
The constraints for this system are 
\be
\chi_1 &\equiv& p_{\sss 2} =0  \\
\chi_2 &\equiv& q^{{\sss 1}2} - q^{{\sss 2}2} + p_{\sss 1}^2 + 
p_{\sss 2}^2 
                             - R^2  =0 
\ee
$$\qquad\qquad\qquad   \eqno(18)	        $$
Constraint analysis is straightforward. The reduced phase space 
is obtained from the inequality
$$ p_{\sss 1}^2 + q^{{\sss 1}2} > R^2                     \eqno(19)   $$ 
$q^{\sss 2}$ has two branches. For each choice
the reduced phase space 
is 2-dimensional infinite plane with a hole of radius $R$ at
the centre, where we have restricted $q^{\sss 2}$ to be positive. 
The reduced Hamiltonian is
$$ H= {{2}\over{3}} \Big( p_{\sss 1}^2 + q^{{\sss 1}2} -R^2 
\Big)^{3/2} 
                                                            \eqno(20)  $$
The system can then be quantized. The Hilbert space consists of 
square-integrable functions on real line ${\bf R}^{\sss 1}$ 
excluding the interval $[-R, R]$.
The evolution will be generated by the Hamiltonian
$$ {\hat H}={{2}\over{3}} \Big( {\hat p}^2 +{\hat q}^2 -R^2
                                \Big)^{3/2}                 $$
As in the earlier case the self-adjointness and positive definiteness of
the Hamiltonian will restrict the Hilbert space.

\section{Conclusion}
We have discussed two singular systems with non-trivial reduced 
phase 
spaces.For the first system the reduced phase space is not $S^2$ 
as
it appears to be but an open disc of radius $R$. If we include the 
boundary  
we can map all points on it to the south pole of $S^2$ with radius
$R/2$. The reduced space would then be $S^2$. However, this can 
be
done only if the singularity at $q^{\sss 2}=0$ is a coordinate
singularity and can be removed by appropriate choice of 
coordinates.
This is not the case for us since the Lagrangian we chose is 
essentially singular at $q^{\sss 2}=0$. For the second system we 
find that the
reduced phase is 2-dimensional plane with a hole at the center. 
This Lagrangian is again singular at $q^{\sss 2}=0$ 
which cannot be removed by any coordinate transformation.  
Canonical quantization reveals restrictions on the Hilbert spaces of 
the two 
systems. These restrictions are manifestations of the classical
constraints at the quantum level.

\acknowledgements{We thank Radhika Vathsan for 
useful discussions and Tabish Qureshi for a critical reading of the 
manuscript.
 One of us (P. Chingangbam) acknowledges financial support 
from Council for Scientific and Industrial Research, India, under 
grant no.9/466(29)/96-EMR-I}

\end{document}